\documentclass [a4paper,aps,showpacs]{revtex4}
\usepackage{graphicx}

\newcommand {\beq}{\equation}
\newcommand {\eeq}{\endequation}
\newcommand {\beqa}{\begin{eqnarray}}
\newcommand {\eeqa}{\end{eqnarray}}
\newcommand {\la} {\langle}
\newcommand {\ra} {\rangle}
\newcommand {\rl} {\rangle\langle}
\newcommand {\inid} {\int\limits_0^\infty d}

\newcommand {\overa} {\hat{\alpha} }

\begin{document}

\title {Zeno era and non-decaying subspaces for multilevel
Friedrichs model}

\author {I.~Antoniou$^{1}$, E.~Karpov$^{1,2}$, G.~Pronko$^{1,3}$, and
E.~Yarevsky$^{1,4}$}

\address{$^1$ Chaos and Innovation Research Unit,
Department of Mathematics, Aristotle University of Thessaloniki,
Thessaloniki, 54124, Greece}
\address{$^2$ International Solvay Institutes for Physics and
Chemistry, Campus Plaine ULB, C.P. 231, Bd. du Triomphe, Brussels
B-1050, Belgium}
\address{$^3$ Institute for High Energy Physics, Protvino, Moscow
region 142284, Russia}
\address{$^4$ Dienst Theoretische Natuurkunde, Vrije Universiteit
Brussel, Pleinlaan 2, B-1050 Brussels, Belgium }

\date {\today}

\begin{abstract}
We study the short-time and medium-time behavior of the survival
probability of decaying states in the framework of the $N$-level
Friedrichs model. The degenerated and nearly degenerated systems
are discussed in detail. We show that in these systems decay can
be considerably slowed down or even stopped by appropriate choice
of initial conditions. We analyze the behaviour of the multilevel
system within the so-called Zeno era. We examine and compare two
different definitions of the Zeno time. We demonstrate that the
Zeno era can be considerably enlarged by proper choice of the
system parameters.
\end{abstract}


\maketitle

\section{Introduction}

Since the very beginning of the quantum mechanics, the measurement
process has been a most fundamental issue which is still not
completely understood. The main characteristic feature of the
quantum measurement is that the measurement changes the dynamical
evolution. This is the main difference of quantum measurement
compared to its classical analogue. In this framework, Misra and
Sudarshan pointed out~\cite{Misra} that repeated measurements can
prevent an unstable system from decaying (the quantum Zeno effect,
QZE).

The QZE has been discussed for many physical
systems~\cite{atom1,atom2,radio,mesoscopic1,mesoscopic2}. Later it
has been found~\cite{antiZeno1,antiZeno2} that under some
conditions the repeated observations could not only slow down but,
in contrary, speed up the decay of the quantum state (the quantum
anti-Zeno effect). The anti-Zeno effect has been further analyzed
in~\cite{PRA61-052107,Nature,AKPY}.

While there exist experiments~\cite{Itano,Toschek} demonstrating
the perturbed evolution of a coherent dynamics, the demonstration
of the QZE for an unstable system with exponential decay, as
originally proposed in~\cite{Misra}, has long been an open
question. Only recently, both Zeno and anti-Zeno effects have been
observed experimentally~\cite{FGMR}.

Zeno effect was recently proposed as a tool to prepare and
maintain the purity of the initial state of quantum
systems~\cite{Zeno-Prepar}, to protect it against
decoherence~\cite{decoherence,Zeno-DecCorr}, and for the
purification of quantum states~\cite{Zeno-Purific}. Multilevel
systems are necessary in order to have decoherence-free subspaces,
as in this case several qubits are required~\cite{PRA63-042307}.
Zeno effect for multilevel systems has already been
discussed~\cite{Zeno-multilevel,Zeno-subspaces}. However, only
non-degenerated systems have been considered so far, although in
actual applications we have quantum systems with (nearly)
identical elements (atoms, ions, quantum dots etc.). Therefore,
the study of the Zeno (and anti-Zeno) effects in (nearly)
degenerated systems is necessary. This is the object of this work
in the frame of the Friedrichs model~\cite{Fried}, which is
appropriate for the discussion of decay and unstable states. The
original Friedrichs model contains two discrete energy levels, a
ground state and an excited state, coupled with the continuum,
which is bounded from below. The time dependence of the survival
probability of excited states has been extensively studied in the
literature~\cite{NPN,P2,AKPY,P3,kofman1,Nature}. For physically
motivated values of the parameters of the model the decay is
approximately exponential with a short non-exponential initial era
and a non-exponential long tail.

The analytical structure of the multilevel Friedrichs model has
also been widely discussed~\cite{Fano,Fn1,Fn2,Fn3,Fn4,Fn5,Exner}.
However, these discussions are mainly concentrated on the
non-degenerated systems, and only in a few papers the time
dependence of the survival probability for $N$-level models has
been discussed~\cite{Davies,Nlevel}.

The degenerated multilevel Friedrichs model is presented in
Section~II. In Section~III we discuss the exponential decay era
and analyze few cases when decay can be substantially slowed down
or even stopped. This discussion is extensively used in the paper.
In Section~IV we analyze the Zeno effect and different Zeno times.
We show that in almost completely degenerate systems, the Zeno era
can be enlarged. In Section~V we analyze the anti-Zeno effect in
the almost completely degenerate systems.

\section{Model and exact solution}

The Hamiltonian of the $N$-level Friedrichs model is:
 \beq \label{H}
  H =  H_0 + \lambda V,
 \eeq
where
\begin{eqnarray}
\label {Ham2}
  H_0 & = & \sum_{k=1}^{N}\omega_k|k\rl k|
        + \int\limits_0^\infty d\omega\,\omega |\omega\rl \omega| ,
        \nonumber \\
V\,\, & = & \sum_{k=1}^N \int\limits_0^\infty d\omega \hat f_k(\omega)
          \left(|k\rl \omega| + |\omega\rl k|\right).
\end{eqnarray}
Here $|k\ra$ represent states of the discrete spectrum with the
energy $\omega_k$, $\omega_k >0$. Degeneracy is reflected in
$\omega_k$. The vectors $|\omega\ra$ represent states of the
continuous spectrum with the energy $\omega$, $\hat f_k(\omega)$
are the form factors for the transitions between the discrete and
the continuous spectrum, and $\lambda$ is the coupling parameter.
The vacuum energy is chosen to be zero. The states $|k\ra$ and
$|\omega\ra$ form a complete orthonormal basis:
\equation \label{Ort}
  \la k| k'\ra   = \delta_{kk'}, \quad
  \la\omega|\omega'\ra = \delta(\omega - \omega'), \quad
  \la\omega|k\ra  = 0, \quad k,k'=1 \ldots N,
\endequation
\beq
\label {Comp}
  \sum_{k=1}^{N} |k\rl k|
  + \int\limits_0^\infty d\omega\,|\omega\rl \omega| = I,
\eeq
where $\delta_{kk'}$ is the Kronecker symbol, $\delta(\omega -
\omega')$ is Dirac's delta function and $I$ is the unity operator.
The Hamiltonian $H_0$ has continuous spectrum over the interval
$[0,\infty)$ and discrete spectrum $\omega_1, ..., \omega_k$
embedded in the continuous spectrum. As the interaction $\lambda
V$ is switched on, the discrete energy levels of $H_0$ become
resonances of $H$ as in the case of the one-level Friedrichs
model~\cite {Fried}. As a result, the total evolution normally
leads to the decay of any initial state in point spectrum
eigenspace:
 \beq \label{init-state}
|\Phi\ra = \sum_k \alpha_k |k\ra, \quad \la \Phi|\Phi \ra = 1.
 \eeq
Decay is described by the survival probability $p(t)$~\cite{Fock}
 \beq \label{survprobdef}
  p(t) \equiv |\langle \Phi|e^{-iHt}|\Phi\rangle|^2 = |A(t)|^2,
 \eeq
where $A(t)$ is the survival amplitude. The survival amplitude can
be calculated as~\cite{Nlevel}
 \beq \label{samp}
  A(t)
    \equiv  \la\Phi|\Phi(t)\ra
    = \sum_{k,m=1}^N \alpha_k \alpha^*_{m}\la k|m\ra_t
    = \sum_{k,m=1}^N \alpha_k \alpha^*_{m} A_{km}(t) \, ,
 \eeq
where the partial survival amplitudes $A_{km}(t)$ are written as
 \beq \label{samp2}
  A_{km}(t)
    = -\sum_j r^j_{km} e^{-iz_j t} + \frac{1}{2\pi i }
    \int_{C}{\rm d}\omega e^{-i\omega t} \hat G_{km}(\omega).
 \eeq
Here $\hat G_{km}(\omega)$ are the matrix elements of the partial
resolvent $\hat G(\omega)$, which is defined as:
\beq
  \hat G^{-1}_{k m}(\omega)
    = (\omega_k - \omega)\delta_{k m} - \lambda^2 \inid \omega'
    \frac{\hat f_k(\omega')\hat f_{m}(\omega')}{\omega' - \omega +i0},
\eeq
The matrix elements $\hat G_{k m}(\omega)$ of the partial
resolvent are analytic in the complex $\omega$ plane with the cut
$[0,+\infty[$. The analytic continuation of $\hat G_{k m}(\omega)$
through the cut has poles $z_j$ at the second Riemann sheet and
$r^j_{km}$ in (\ref{samp2}) is the residue of $\hat
G_{km}(\omega)$ at the pole $z_j$. The contour of integration $C$
in (\ref{samp2}) runs from $-i\infty$ to 0 on the first sheet of
the Riemann surface and then from 0 to $-i\infty$ on the second
sheet.

In the following, it will be more convenient to consider the
dimensionless variable $x=\omega/\Lambda$, where $\Lambda$ is a
parameter with the dimension of energy. Due to the dimension
argument, we can write the form factor $\hat f(\omega)$ in the
form
$$
\hat f(\omega) = \sqrt{\Lambda} f(x),
$$
where $f(x)$ is a dimensionless function. Introducing the
dimensionless partial resolvent $G(x) = \Lambda \hat G(\omega)$,
we find
 \beq \label{eta}
  G^{-1}_{k m}(x)
    = (x_k - x)\delta_{k m} - \lambda^2 \inid x'
      \frac{f_k(x') f_{m}(x')}{x' - x +i 0},
 \eeq
and
\beq
  A_{km}(t)
    = -\sum_j r^j_{km} e^{-i \Lambda x_j t} + \frac{1}{2\pi i }
    \int_{C}{\rm d}x e^{-i \Lambda x t} G_{km}(x).
\eeq

\section{The exponential era}

We concentrate mainly on the model close to the completely
degenerate one while the model without degeneracy has been
discussed previously~\cite{Nlevel}. We analyze the behaviour of
the system in the exponential era (i.e. the poles structures
mainly) for identical (A) and different (B) form factors.

We suppose that the form factors can be expressed as
 \beq
f_k(x) = p^{k-1} f(x) + \varepsilon q_k(x), \quad \mbox {$p$ \ is
a constant.}
 \eeq
We assume that $\varepsilon$ is small, and consider the expansion
in the vicinity of $\varepsilon=0$. This choice is motivated by
the expected similarity of the form factors for the degenerate
levels. Then the integrals in Eq.~(\ref{eta}) can be written as
$$
\int_0^\infty dx' {f_k(x')f_m(x') \over x'-x+i0} = W_{km}(x) +
\varepsilon R_{km}(x),
$$
where
 \beqa
W_{km}(x) = p^{k+m-2} W(x)= p^{k+m-2}\int_0^\infty dx' {f^2(x')
\over x'-x+i0}, \\
 R_{km}(x)=\int_0^\infty dx' {f(x')(p^{k-1} q_m(x')+ p^{m-1}
q_k(x')) \over x'-x+i0}+ \varepsilon \int_0^\infty dx'
{q_k(x')q_m(x') \over x'-x+i0}.
 \eeqa
The matrix elements of the inverse partial resolvent~(\ref{eta})
are:
 \beq \label{resolv}
G_{km}^{-1}(x) = (x_k -x)\delta_{km} - \lambda^2 p^{k+m-2} W(x) -
\lambda^2 \varepsilon R_{km}(x).
 \eeq
As both matrix elements $W(x)$ and $R(x)$ are complex, the
resolvent $G(x)$ is also a complex function.

\subsection{Identical form factors}

For $\varepsilon=0$, the partial resolvent can be found
explicitly:
 \beq \label{exact-res}
G_{km}(x) = {\delta_{km} \over x_m - x} + {\lambda^2 p^{k+m-2}
W(x) \over (x_m - x) (x_k - x) (1 - \lambda^2W(x)\sum_i p^{2(i-1)}
(x_i - x)^{-1})},
 \eeq
as well as the determinant
 \beq \label{exact-det}
\mbox {det}G^{-1}(x) = (1-\lambda^2W(x) \sum_i
p^{2(i-1)}(x_i-x)^{-1})\prod_k(x_k-x).
 \eeq
When the energy levels $x_k$ are well separated, each of them
becomes a resonance $z_k$ for nonzero $\lambda^2$:
$$
z_k = x_k - \lambda^2 p^{2(k-1)} W(x_k) + O(\lambda^4).
$$
This case is discussed in detail in~\cite {Nlevel}.

As we focus on the model close to the completely degenerate one,
we suppose that $p=1$. First, we consider the case when all energy
levels are degenerated: $x_k=\bar x$ for any $k$. Then we deduce
the partial resolvent and the determinant:
 \beqa \label{GMexact}
G_{km}(x) = {1 \over \bar x - x} \left(\delta_{km} + {\lambda^2
W(x) \over \bar x - x - N \lambda^2 W(x)}\right), \\ \mbox
{det}G^{-1}(x) = (\bar x-x)^{N-1}(\bar x-x-N\lambda^2W(x)).
 \eeqa

The matrix elements~(\ref{GMexact}) have two poles in the vicinity
of $\bar x$: the real pole $z_1=\bar x$ and the complex pole $z_2$
defined by the equation $\bar x - z_2 - N \lambda^2W(z_2)=0$. For
small $\lambda^2$ we find
 \beq \label{case1-z2}
z_2 = \bar x - \lambda^2 N W(\bar x) + O(\lambda^4).
 \eeq
The residues of $G_{km}(x)$ at the poles are
\begin{eqnarray} \label{eq33}
 \mbox {res}_{x=z_1}G_{km}(x) = -\delta_{km} +{1 \over N},
 \nonumber \\
\mbox {res}_{x=z_2}G_{km}(x) = {-1 \over N(1 + \lambda^2 N
W'(z_2))} = {-1 \over N(1 + \lambda^2 N W'(\bar x))} +
O(\lambda^4),
\end{eqnarray}
where $W'(x)$ is the derivative of $W(x)$.

Let us now find the time evolution of state
$|\Phi\ra$~(\ref{init-state}). In our case, there exist only two
poles $z_1$ and $z_2$ in a vicinity of the real axis. The survival
amplitude $A(t)$ is
 \beq \label{ampl-deg}
A(t) = e^{-i\bar{x} \Lambda t} \left( 1 - {1 \over N} (1 -
{e^{-i(z_2-\bar{x})\Lambda t} \over 1 + \lambda^2 N W'(z_2) })
|\overa|^2 \right) +
 {\lambda^2 |\overa|^2 \over 2\pi i} \int_C dx
 {W(x) e^{-ix\Lambda t} \over (\bar x - x - N \lambda^2 W(x))
 (\bar x - x)},
 \eeq
where we have introduced
 $$ \overa = \sum_k \alpha_k.$$
We observe that for almost all initial states, the oscillations of
the survival amplitude may not vanish with time. Indeed, if
$\overa=0$ then there is no decay at all and the survival
probability is $p(t)=|A(t)|^2=1$. The survival probability decays
to zero if and only if the initial state is
$\alpha_k=e^{i\phi}/\sqrt{N}$ for any $k$ and some real $\phi$.
For such states one has $|\overa|^2=N$. For an arbitrary initial
state, the decay is incomplete and
 $$\lim\limits_{t\to\infty} p(t) =
 \left(1-{|\overa|^2 \over N}\right)^2 \neq 0. $$
We note that the decay defined by Eq.~(\ref{ampl-deg}) is
oscillating~\cite{Nlevel}. The frequency of oscillations is
 $$F_{osc}= \mbox{Re}(z_2-\bar{x})\Lambda/2\pi \sim \lambda^2 N
 \Lambda. $$
The decay rate is
 $$\gamma = \mbox{Im}(z_2) \Lambda \sim \lambda^2 N
 \bar x \Lambda,$$
which is considerably smaller then $F_{osc}$ for the physically
motivated form factors. This oscillating decay is presented in
Fig.~1. However, if the parameters of the quantum system are such
that the system experiences complete decay (i.e.
$\lim\limits_{t\to\infty} p(t) = 0$) then there are no
oscillations in the exponential term  in (\ref{ampl-deg}). The
integral term is negligible in the exponential era, therefore the
system decays without oscillations, that is also shown in Fig.~1.

\begin{figure}[hbt]
 \centering \includegraphics[width=0.6\textwidth]{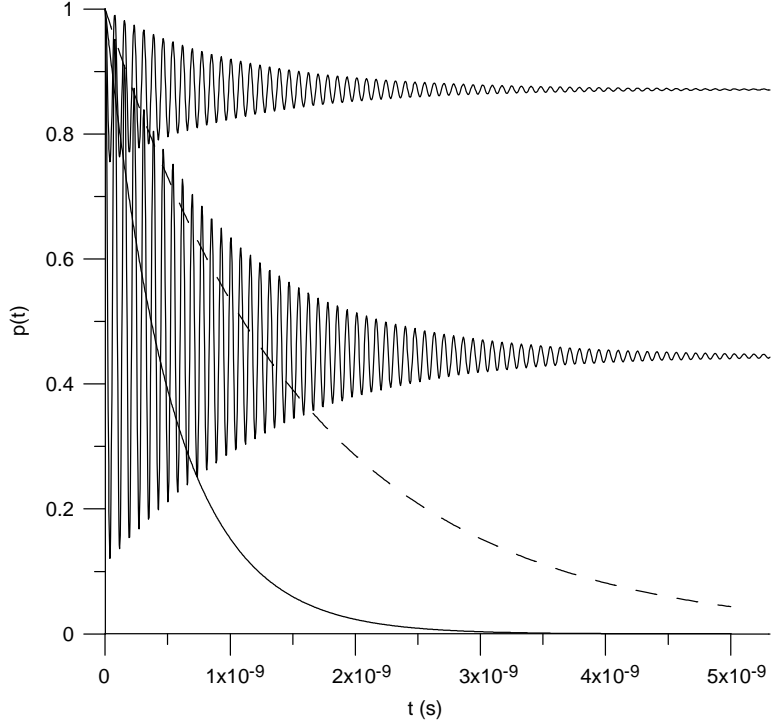}
\caption{The survival probability for the completely degenerated
system. From above, the curves correspond to $N=3$, $|\overa|^2$ =
0.2, 1.0, 3.0. For comparison, the decay of the one-level system
is also shown (the dashed line). The parameters of the model are
selected for the hydrogen atom: $\Lambda=8.498 \ 10^{18}$s$^{-1}$,
$\omega= 1.55 \ 10^{16}$s$^{-1}$, $\lambda^2=6.43 \ 10^{-9}$.}
\end{figure}

Let us now discuss the situation when the system is not completely
degenerate, but it is close to the degenerate one. Namely, we
consider the case when one energy level differs from others:
$x_k=\bar x$ for $k=1\ldots N-1$, $x_N=\bar x +\Delta$. If the
form factors are identical, the determinant~(\ref{exact-det})
takes the following form:
 $$
 \mbox {det}G^{-1}(x) = (\bar x-x)^{N-1}(x_N-x)-
 \lambda^2 W(x) ((N-1)(\bar x-x)^{N-2}(x_N-x)+(\bar x-x)^{N-1}).
 $$
It has three different roots: the root $z_1 = \bar x$ with the
multiplicity $N-2$ and the roots $z_{2,3}$ defined by
 \beq \label{z23full}
 z_{2,3} = \bar x +{\Delta -\lambda^2 N W(z)
 \mp \sqrt{(\Delta -\lambda^2 N W(z))^2 + 4\lambda^2 \Delta (N-1)
 W(z)} \over 2} .
 \eeq
Expression~(\ref{z23full}) gives the values for the roots for any
$\Delta$ and $\lambda^2$. However, the limit when both these
parameters go to zero is irregular. We will show that the pole and
resolvent structure depends on the order, in which the limits are
taken.

When $\lambda^2 \ \mbox{Re } W(\bar x) \ll \Delta$, we find
 \beq
 z_2 = \bar x - \lambda^2 (N-1) W(\bar x) + O(\lambda^4), \quad
 z_3 = x_N - \lambda^2 W(x_N) + O(\lambda^4).
 \eeq
The root $z_2$ corresponds to the root~(\ref{case1-z2}) (the
multiplicity is less by 1), and the root $z_3$ in this case is
well-separated from $z_{1,2}$. Therefore the structure of the
residues of the partial resolvent is composed of two independent
blocks: one block coincides with Eq.~(\ref{eq33}) for the
multiplicity $N-1$, and the second block corresponds to the
non-degenerated level (i.e. $\mbox {res}_{x=z_3}G_{NN}(x)=-1$, and
all other residues are equal to 0).

For the situation when $\Delta \ll \lambda^2 \ \mbox{Re } W(\bar
x)$, we find
 \beq \label{z23delta}
 z_2 = \bar x - \lambda^2 N W(\bar x) + O(\Delta), \quad
z_3 = x_N - {1 \over N}\Delta + {\Delta^2 \over \lambda^2 W(\bar
x)} {N-1 \over N^3} + O(\Delta^3).
 \eeq
Again, the root $z_2$ corresponds to the root~(\ref{case1-z2}) and
its imaginary part does not disappear when $\Delta\to 0$. The
third root $z_3$ becomes real for identical energies, and the
corresponding decay rate
 $$\gamma_3 = -{2\pi \Lambda f^2(\bar x)
 \Delta^2 \over \lambda^2 |W(\bar x)|^2}{N-1 \over N^3}
 \to 0 \quad \mbox{when} \quad \Delta\to 0. $$
The residues of the resolvent $G(x)$ at poles of the determinant
are
 $$
\mbox{res}_{x=z_1}G(x) =
 \left( \begin{array}{cc}
- I^{(N-1)}+{1 \over N-1}P^{(N-1)} & 0 \\ 0 & 0
 \end{array} \right ), \quad
\mbox {res}_{x=z_2}G(x) = - {1 \over N} P^{(N)},
 $$
 \beq
\mbox {res}_{x=z_3}G(x) =
 \left( \begin{array}{cc}
{1 \over N(1-N)} P^{(N-1)} & {1 \over N} \\ {1 \over N} & {1 \over
N}-1
 \end{array} \right),
 \eeq
where $I^{(m)}$ is the $m \times m$ unit matrix and the $m \times m$
matrix $P^{(m)}$ is defined as $P^{(m)}_{ik}=1$.
The survival amplitude $A(t)$ is
 \beqa \label{A-1diff}
A(t) & \approx & e^{-i \bar{x} \Lambda t} \left( \sum_{k=1}^{N-1}
|\alpha_k|^2 - {1 \over N-1}|\sum_{k=1}^{N-1} \alpha_k|^2 + {1
\over N} e^{-i(z_2-\bar{x})\Lambda t} |\overa|^2 \right. \nonumber
\\
 & + & \left. e^{-i(z_3-\bar{x})\Lambda t}
\left({1 \over N-1}|\sum_{k=1}^{N-1} \alpha_k|^2 + |\alpha_N|^2 -
{1 \over N} |\overa|^2 \right) \right).
 \eeqa

This result reproduces formula (\ref{ampl-deg}) when $\Delta\to
0$. In our assumptions, we have two time scales for exponential
decay: a fast decay defined by $z_2$ with decay rate proportional
to $\lambda^2 N$, and a slow decay defined by $z_3$ with decay
rate proportional to $\Delta^2$. This slow decay is manifestation
of non-degeneracy. The non-decaying subspaces of the system are
now defined by two conditions: $|\overa|=0$ and $\alpha_N=0$.

\subsection{Different form factors}

In the case of different form factors, $\varepsilon \neq 0$, one
cannot find a general explicit expression for the matrix elements
of the partial resolvent $G_{km}$ and its determinant. However, as
the problem is the eigenvalue problem for a finite matrix, a
general qualitative description is known (see e.g. theorem~XII.2
in~\cite{Reed4}). According to this theorem, for the system with
identical energies $x_k \equiv \bar x$ in the vicinity of $\bar x$
there exist $N-1$ roots of the determinant, additionally to the
root $z_2$ (\ref{case1-z2}). Generally speaking, these roots give
rise to exponentially decreasing terms in the survival amplitude
$A(t)$. However, there may also exist real roots corresponding to
bound states. These roots result in non-decaying behaviour of the
survival probability. The existence and the number of such roots
depend on specific properties of the form factors.
%%For example, if the form factors $f_k(x)$ are linearly dependent
%%for some $\varepsilon$, one of these roots coincides with $\bar
%%x$.

Having in mind this qualitative description we shall analyze the
poles structure of $G(x)$ by perturbation expansion. In the first
non-vanishing order of the perturbation expansion with respect to
$\varepsilon$, we have for the roots of the determinant the
following equation:
 \beq
\det G_0^{-1}(x) - \varepsilon \lambda^2 \det G_0^{-1}(x) \mbox
{tr} \left( G_0(x) R(x)\right) = 0,
 \eeq
where the resolvent is defined by Eq.~(\ref{resolv}): $G_0(x)
\equiv G|_{\varepsilon=0}(x)$. This equation gives us the
following roots: $N-2$ roots of the type of $z_1=\bar{x}$, the
root $z_2$ (\ref{case1-z2}) and the new root $z_3$:
 \beq
z_3 = \bar x - \varepsilon^2 \lambda^2 \mbox {tr} \left( (I -
{1\over N} P) Q(\bar x) \right) + O(\varepsilon^3 \lambda^2 ),
 \eeq
where the $N \times N$ matrix $Q(x)$ is defined as
 $$Q_{ik}(x)=\int dx' {q_i(x')q_k(x') \over (x'-x+i0)}.$$
We note that the perturbation series for $z_3$ starts from the
$\varepsilon^2$-term (see also a unified framework for noiseless
quantum subsystems in~\cite{PRA63-042307}), while for one
non-degenerated level~(\ref{z23delta}) $\mbox{Re} (z_3-\bar x)
\sim \Delta$ and $\mbox{Im} z_3 \sim \Delta^2$. In the present
case, the imaginary part of $z_3$ can be calculated as
$$
\mbox{Im} z_3 = 2\pi \varepsilon^2 \lambda^2 \left( \sum_k
q_k^2(\bar x) -{1 \over N} \left(\sum_k q_k (\bar x)\right)^2
\right)+ O(\varepsilon^3 \lambda^2 ).
$$
One can see that $\mbox{Im} z_3$ can be equal to zero even for
different form factors $q_k(x)$. In this case, the decay will be
slower, its width will be proportional to $\varepsilon^3
\lambda^2$.

When one considers next terms of the perturbation series with
respect to $\varepsilon$, more resonances are split from the
energy level $\bar x$. Generally speaking, each additional term in
the perturbation series gives an additional resonance so
$\mbox{Im} z_k \sim \varepsilon^{k-1}$ for $k=3\ldots N$, and all
the roots become complex when the complete series is taken into
account.

\section{Zeno effect and Zeno time}

In order to estimate the short-time behaviour for the
system~(\ref{Ham2}), we will use the Taylor expansion of the
survival probability. We shall assume here the existence of all
necessary matrix elements, and denote the expectation values as
$\la \cdot \ra = \la \Phi |\cdot| \Phi \ra$. Then, following the
results of~\cite{AKPY2003}, we find
\begin{eqnarray} \label{p-expansion}
 p(t) = \la e^{-iHt} \ra =
  1-t^2 \left(\la H^2 \ra - \la H \ra^2\right) +
t^4 \left({1\over 4} \la H^2 \ra^2 +{1\over 12} \la H^4 \ra -
 {1\over 3} \la H \ra \la H^3 \ra \right) +O(t^6) =  \nonumber \\
 1 - {t^2\over t_a^2} + {t^4\over t_b^4} + O(t^6).
\end{eqnarray}

The expressions for the times $t_a$ and $t_b$ can be deduced using
the special structure of the potential $V$~(\ref{Ham2}):
 \beqa \label{ta}
 {1 \over \Lambda^2 t_a^2} = \sum_k |\alpha_k|^2 x_k^2 -
\left( \sum_k |\alpha_k|^2 x_k\right)^2 + \lambda^2 R_1, \\
 {1 \over \Lambda^4 t_b^4} = {1 \over 4}
 \left(\sum_k |\alpha_k|^2 x_k^2 \right)^2
+ {1 \over 12} \sum_k |\alpha_k|^2 x_k^4 - {1 \over 3} \sum_k
|\alpha_k|^2 x_k \sum_k |\alpha_k|^2 x_k^3 \nonumber \\
 + \lambda^2 \left( {1 \over 2} R_1 \sum_k |\alpha_k|^2
x_k^2 + {1 \over 12} R_3 - {1 \over 3} R_2 \sum_k |\alpha_k|^2 x_k
\right)
 + \lambda^4  \left( {1 \over 12} R_4 + {1 \over 4} R_1^2
 \right).
 \label{tb}
 \eeqa
Here
 \beqa
 R_1 & = & \sum_{ik} \alpha_i\alpha_k^* F_{ik}^0, \\
 R_2 & = & \sum_{ik} \alpha_i\alpha_k^*
\left( (x_i+ x_k)F_{ik}^0 + F_{ik}^1 \right), \\
 R_3 & = & \sum_{ik} \alpha_i\alpha_k^*
 \left( (x_i^2+ x_i x_k+ x_k^2)F_{ik}^0 +
 (x_i+x_k) F_{ik}^1 + F_{ik}^2 \right), \\
 R_4 & = & \sum_{ik} \alpha_i\alpha_k^* ((F^0)^2)_{im} ,
 \eeqa
where
 $$
F_{ik}^p = \int_0^\infty dx \, x^p f_i(x) f_k(x).
 $$

The probability that the state $|\Phi\ra$ after $M$ equally spaced
measurements during the time interval $[0,T]$ has not decayed, is
$p_M(T) = p^M(T/M)$~\cite{Misra}. We are interested in the limit
$M \rightarrow \infty$ or, equally, the time interval between the
measurements $\tau=T/M$ goes to zero:
\begin{equation} \label{Zenoas}
\lim\limits_{\tau \rightarrow 0}
 p_M(T) = \lim\limits_{\tau \rightarrow 0} p(\tau)^{T/\tau}
       = \left\{\begin{array} {ll}
                          1, & {\rm when} \quad  p'(0) = 0, \\
                    e^{-cT}, & {\rm when} \quad p'(0) = -c, \\
                          0, & {\rm when} \quad p'(0) = -\infty.
              \end{array}  \right.
\end{equation}
This limit corresponds to so-called continuously ongoing
measurements during the entire time interval $[0,T]$. Obviously,
this is an idealization. In practice we have a manifestation of
the Zeno effect if the probability $p_M(T)$ increases as the time
interval $\tau$ between measurements decreases.
Formula~(\ref{Zenoas}) may be accepted as an approximation only
for very short time intervals $\tau$.

As one refers in discussions about the Zeno effect on the Taylor
expansion~(\ref{p-expansion}) of survival probability for short
times, and specifically on the second term, we shall define the
Zeno time $t_Z$ as corresponding to the region where the second
term dominates. Hence, one way is to define the Zeno time $t_Z$
(see paper~\cite {AKPY}) as a boundary where the second and third
terms have the same amplitude:
\begin{equation}\label{zenotime}
{t_Z^2 \over t_a^2} = {t_Z^4 \over t_b^4}, \quad \mbox {so} \quad
t_Z=t_b^2/t_a.
\end{equation}

Another way to discuss the Zeno and the anti-Zeno effects is the
variable decay rate $\gamma(t)$~\cite{Pascazio-var}. Namely, we
represent the survival probability as $p(t)=e^{-2\gamma(t) t}$.
Then we find
 \beq
 \gamma(t) = - {1\over 2t} \log{p(t)}.
 \eeq
In terms of the $\gamma(t)$, we find that
$p_M(T)=e^{-2\gamma(T/M)\,T}$. Therefore, we have the Zeno region
(deceleration of the decay) when $\gamma(\tau) < \gamma(T)$ and
the anti-Zeno region (acceleration of the decay) when
$\gamma(\tau) > \gamma(T)$. If the expansion~(\ref{p-expansion})
is valid, then $\gamma(\tau) \approx \tau/(2t_a^2)$ for $\tau
\lesssim t_Z$, and we always have the Zeno region for short times.
Then we can define the Zeno time $\tau_Z$ as the boundary between
the Zeno and anti-Zeno regions: $\gamma(\tau_Z)=\gamma(T)$. Again,
if the expansion~(\ref{p-expansion}) is valid and $T$ is within
the exponential era, then $\tau_Z \approx 2t_a^2/t_d$, where $t_d$
is the time interval when the decay is almost exponential (the
so-called ``exponential era'').

Even if expansion~(\ref{p-expansion}) is not valid, we can still
apply the same idea and use the short-time expansion of the
survival probability. For example, for the form factor $\sqrt{x}
\over 1 + x $ one finds in the one-level Friedrichs
model~\cite{AKPY} that $p(t) \approx 1-(t/t_a)^{1.5}$ for small
$t$, and $\tau_Z \approx 4t_a^3/t_d^2$. We compare two definitions
of the Zeno time in Table~1 for different form factors in the
frame of the one-level Friedrichs model~\cite{AKPY}. One can see
that $\tau_Z \approx C t_Z {\omega_1 \over \Lambda} \ll t_Z$ for
physically motivated parameters. Hence the Zeno time $\tau_Z$ is
even shorter than the previously estimated time $t_Z$~\cite{AKPY}.
The time $\tau_Z$ gives an interesting connection between the
decay time and the energy uncertainty of the initial state (as
$t_a^{-2} = \la H^2 \ra - \la H \ra^2$ when these matrix elements
exist).

\begin{table}
\caption{The Zeno times $t_Z$ and $\tau_Z$, the time $t_a$ and the
decay time $t_d$ for two model of interactions in the one-level
Friedrichs model.}
\begin{tabular}{ccccc} \hline
 Form factor & $t_a$ & $t_d$ & $t_Z$ & $\tau_Z$ \\ \hline
 $\sqrt{x} \over 1 + x $ &
 $({3\over 4\sqrt{2\pi}})^{2/3}\over \lambda^{4/3}\Lambda$ &
 $ {1\over \pi\lambda^2 \sqrt{\Lambda \omega_1}}$ &
 ${32\over 9\pi}{1\over\Lambda}$ &
 ${9\pi\over 8} {\omega_1\over \Lambda^2}$ \\
 $x \over \left(1 + x^2 \right)^4 $ & $\sqrt{6} \over \lambda\Lambda$
 & ${1\over 2\pi\lambda^2 \omega_1}$ & $ 2\sqrt{6} \over \Lambda$ &
 ${24\pi\omega_1\over \Lambda^2}$ \\\hline
\end{tabular}
\end{table}

Expressions (\ref{ta},\ref{tb}) include many different parameters
and can hardly be analyzed in the general case. We consider here
three specific representative examples for the physically
motivated weak coupling model~\cite{AKPY} with
\begin{equation} \label{labelA}
\lambda^2 \ll 1 \quad \mbox{and} \quad x_k = \omega_k /\Lambda \ll
1.
\end{equation}

{\bf Example A. The decay of one level}

We consider here the multilevel Friedrichs model. We do not
introduce any assumption on the energy levels and the form
factors. The initial condition was chosen so that the only one
level $l$ is occupied: $\alpha_l=1$, $\alpha_k=0$ for $k \neq l$.
Then expressions (\ref{ta},\ref{tb}) become
\begin{eqnarray}
{1 \over t_a^2} &=& \lambda^2 \Lambda^2 F_{ll}^0, \nonumber \\
 {1 \over t_b^4} &=& \lambda^2 \Lambda^4 \left( {1 \over 12}
x_l^2 F_{ll}^0 - {1 \over 6} x_l F_{ll}^1 + {1 \over 12} F_{ll}^2
\right) + \lambda^4 \Lambda^4 \left( {(F_{ll}^0)^2 \over 4} + {1
\over 12} ((F^0)^2)_{ll} \right). \nonumber
\end{eqnarray}
It is not surprising that the expressions for $t_a$ and $t_b$
practically coincide with those for the one-level Friedrichs
model~\cite{AKPY}. Therefore, the Zeno time $t_Z$ is also the
same:
$$
t_Z \sim {1\over \Lambda} \sqrt{12 F^0_{ll}\over F^2_{ll}}.
$$
The only difference in the last term for $t_b^4$ ($\sum_m F_{lm}^0
F_{ml}^0$ instead of $(F_{ll}^0)^2$ for one-level model) does not
influence the results for a number of levels $N \ll 1/\lambda^2$
and $\lambda^2 \ll 1$.

The Zeno time $\tau_Z$ is more sensitive to the structure of the
energy levels than $t_Z$. If the non-degenerated energy level
$x_l$ is well separated from the others, then the survival
probability in the exponential era coincides with the survival
probability for the one-level Friedrichs model. In this case, the
decay time $t_d=1/ \mbox{Im} z_l$, and $\tau_Z = t_a^2/t_d$ as for
the one-level model. As we shall show later, the situation is
quite different when the energy level is degenerated.

{\bf Example B. The completely degenerate case}

In this case, all frequencies are identical, $x_k= \bar x$ for any
$k$. Expressions (\ref{ta},\ref{tb}) become
 \beqa
 {1 \over t_a^2} &=& \lambda^2 \Lambda^2 \la F^0 \ra, \nonumber \\
{1 \over t_b^4} &=& \lambda^2 \Lambda^4 \left( {1 \over 12} {\bar
x}^2 \la F^0 \ra - {1 \over 6} {\bar x} \la F^1 \ra + {1 \over 12}
\la F^2 \ra \right) + \lambda^4 \Lambda^4 \left( { \la F^0 \ra ^2
\over 4} + {1 \over 12} \la (F^0)^2 \ra \right), \nonumber
 \eeqa
where
$$
\la F^p \ra = \sum_{ik} \alpha_i\alpha_k^* F_{ik}^p .
$$
The matrices $F^p$ are the Gramm matrices, which have the
following property~\cite{Gramm}:
 \begin{eqnarray} \label{gramm}
\la F^p \ra & \geq & 0, \\
 \la F^p \ra & = & 0 \ \ \mbox{for
some $|\tilde\Phi\ra$, iff the form factors $f_k(x)$ are linearly
dependent:} \ \sum_k m_k f_k(x) \equiv 0. \nonumber
 \end{eqnarray}
When all averages $\la \cdot \ra $ are separated from zero, e.g.
the form factors $f_k(x)$ are linearly independent, it resembles
Example~A. The Zeno time is of the same order of magnitude
$$
t_Z \sim {1\over \Lambda} \sqrt{12\la F^0 \ra \over \la F^2 \ra }.
$$
However, in the case of linearly dependent form factors,
the situation may change for special initial conditions $|\Phi\ra$.
Assuming for the sake of simplicity that the form factors are
identical, $\bar x \ll 1$, $\lambda^2 \ll 1$, we find that
\begin{equation}
{1 \over t_a^2} = \lambda^2 \Lambda^2 F^0 |\overa|^2, \quad
 {1\over t_b^4} \approx {\lambda^2 \Lambda^4 \over 12} F^2
|\overa|^2 , \quad t_Z \approx {1 \over \Lambda} \sqrt{12 F^0
\over F^2}.
\end{equation}
We can now see that the Zeno time $t_Z$ is independent of the
initial conditions. However, both $t_a$ and $t_b$ increase to the
infinity when $\overa$ goes to zero. This means that the state
$|\Phi\ra$ with $\overa=0$ does not decay, while the relation
between $t_a$ and $t_b$ is unchanged. The same is true when the
form factors are not identical but linearly dependent: there also
exists a non-decaying state $|\tilde\Phi\ra$~(\ref{gramm}).

The Zeno time $\tau_Z$ essentially depends on the time of the
observation $T$. If $T$ is considerably bigger than the decay time
$t_d$, then $\gamma(T) \approx -\log{(1-|\overa|^2/N)}/T$, and the
Zeno time
 $$
 \tau_Z \approx {-2\log{(1-|\overa|^2/N)} \over
 \lambda^2\Lambda^2 F^0 |\overa|^2 T}.
 $$
If the observation time $T \sim C t_d$, then for small
$|\overa|^2$ we find $\tau_Z \sim {\bar x}/\Lambda$. Hence both
times $t_Z$ and $\tau_Z$ have for small $|\overa|^2$ finite values
while the initial state $|\Phi\ra$ does not decay for
$|\overa|^2=0$.

Therefore, both Zeno times $t_Z$ and $\tau_Z$ may not be relevant
for the $N$-level model and should be modified. We would like to
note that this problem does not appear for the one-level model
analyzed in~\cite{AKPY}.

{\bf Example C. $N$-level model with one different level}

In this case, the energy of one level differs from the others:
$x_k= \bar x $, $k=1\ldots N-1$, $x_N= \bar x + \Delta$, and the
form factors are identical: $f_k(x)=f(x)$. Then we have for the
time $t_a$
$$
{1 \over t_a^2} = \Lambda^2\Delta^2 (|\alpha_N|^2-|\alpha_N|^4) +
\lambda^2 \Lambda^2 F^0 |\overa|^2.
$$
The time $t_b$ is first analyzed under conditions~(\ref{labelA}):
$$
{1 \over t_b^4} \approx {\Lambda^4\Delta^4 \over 12}
(|\alpha_N|^2-|\alpha_N|^4) + {\lambda^2 \Lambda^4 \over 12} F^2
|\overa|^2.
$$
Both these times and the Zeno times depend on the initial vector.
One can easily see that if $|\overa|^2 \neq 0$, the Zeno time
$t_Z$ has a maximum as the function of the energy difference
$\Delta$. We can estimate the position and the value of this
maximum:
\begin{equation} \label{Zeno-time-max}
t_Z \left(\Delta \approx \sqrt{\lambda} \sqrt[4]{F^2 |\overa|^2
\over |\alpha_N|^2-|\alpha_N|^4 }\right) \approx \sqrt[4]{36
(|\alpha_N|^2-|\alpha_N|^4) \over F^2 |\overa|^2} {1 \over
\Lambda\sqrt{\lambda}}.
\end{equation}
We observe that the Zeno time is increased by the factor $\sim
1/\sqrt{\lambda}$ with respect to the one-level model. We note
that even if we consider the exact expression for $t_b$, the same
increase of the Zeno time takes place. We illustrate this increase
on Fig.~2a using exact expression~(\ref{tb}) for the time $t_b$.
We use the parameters of the model associated with the hydrogen
atom~\cite{P1}. One can see that for this system the
maximum~(\ref{Zeno-time-max}) cannot be reached. However, for
small $\Delta$ we can find
\begin{equation}
t_Z \left(\Delta \right) \approx t_Z(0) \left ( 1+
{|\alpha_N|^2-|\alpha_N|^4 \over 2\lambda^2 F^0 |\overa|^2}
\Delta^2 \right ) > t_Z(0),
\end{equation}
therefore some increase of the Zeno time always takes place.

\begin{figure}[hbt]
 \centering \includegraphics[width=0.45\textwidth]{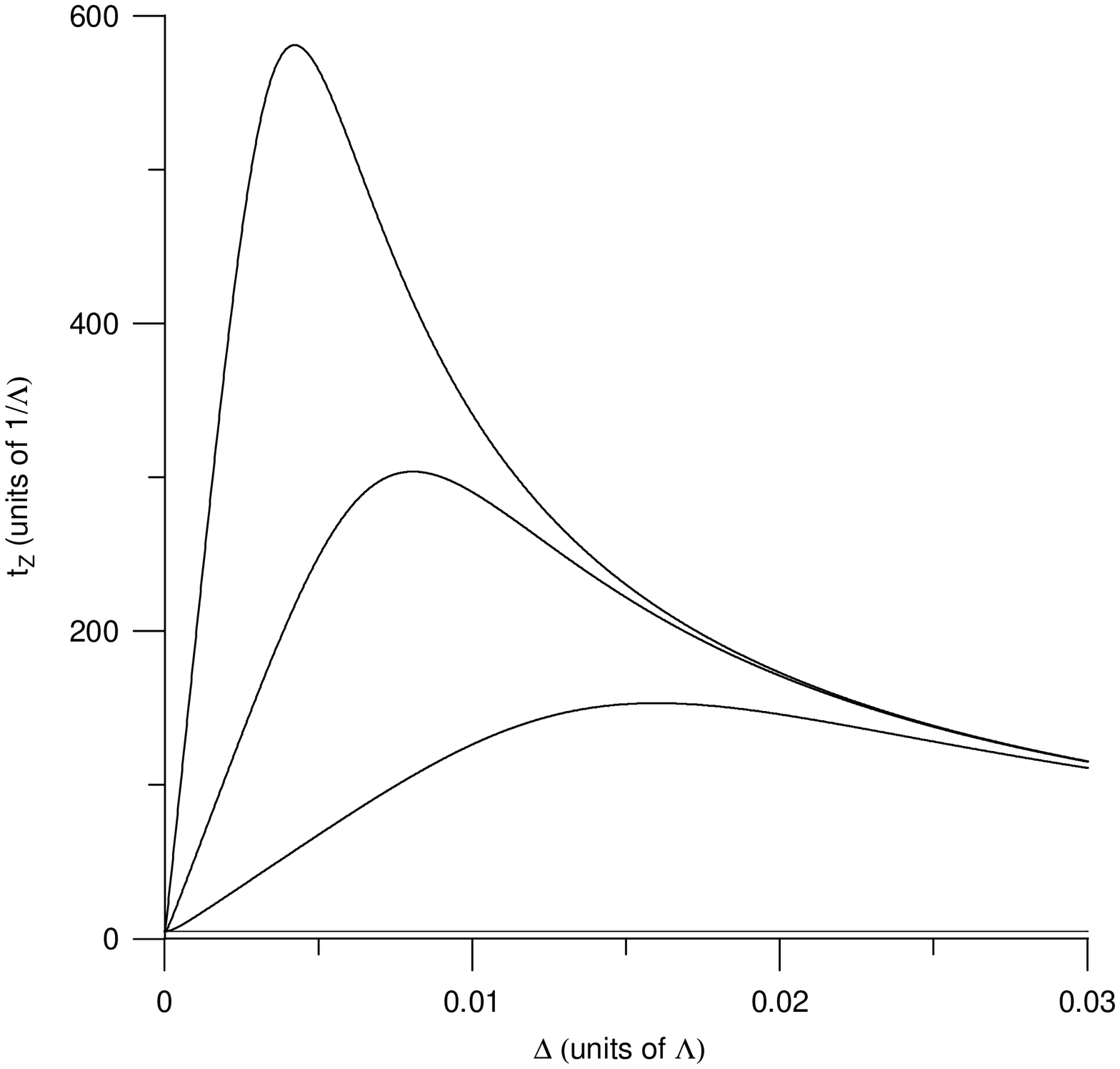}
 \centering \includegraphics[width=0.45\textwidth]{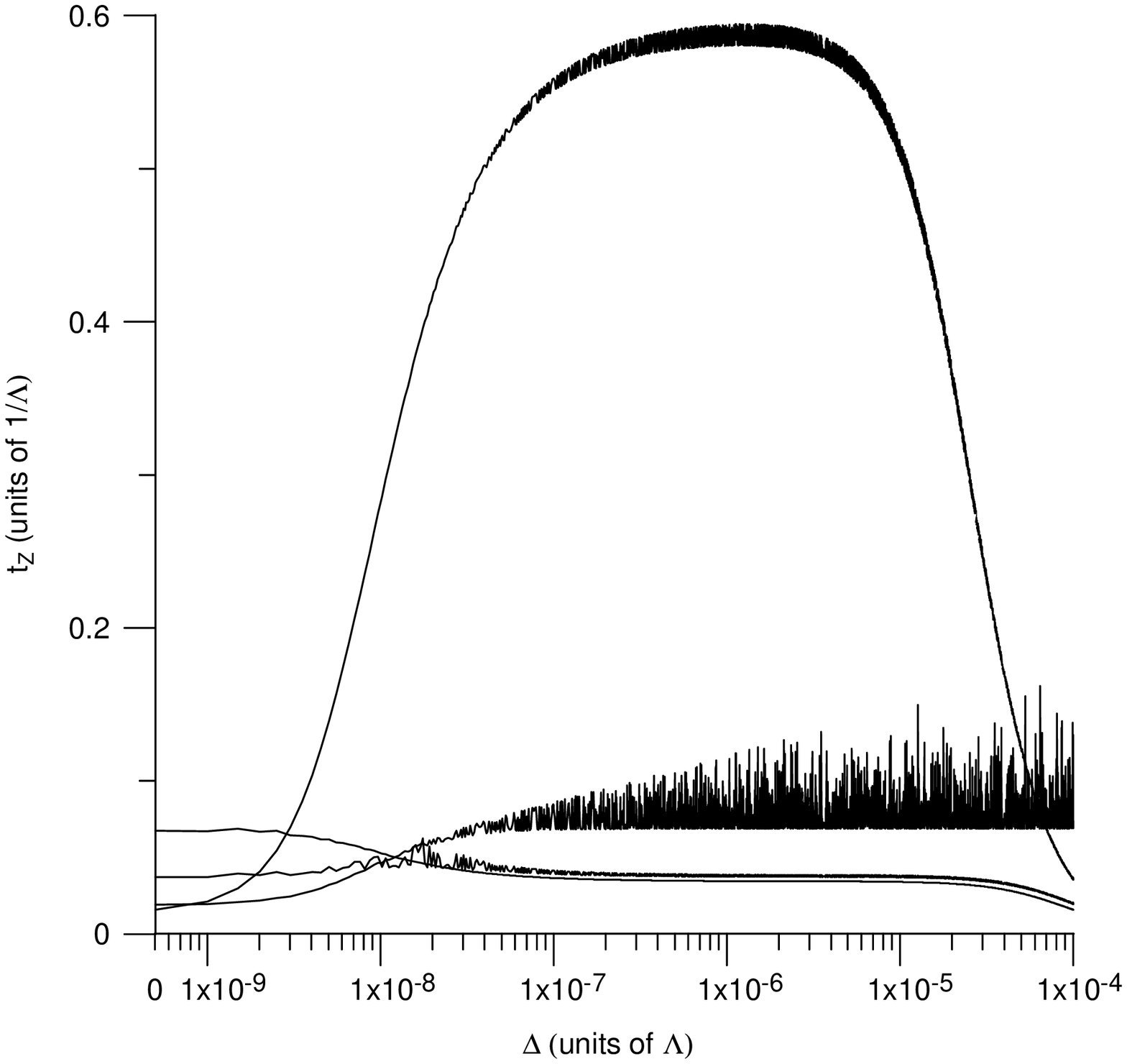}
\caption{The Zeno time $t_Z$ (Fig.~2a) and the Zeno time $\tau_Z$
(Fig.~2b) as a function of the energy difference $\Delta$ for
two-level Friedrichs model.  From above, the curves correspond to
the initial condition $|\Phi\ra$: (2a)
$(\alpha_1,\alpha_2)=(1,-0.6)$, $(1,1)$, $(1,0.1)$ and $(1,0)$;
(2b) $(\alpha_1,\alpha_2)=(1,0.95)$, $(1,0.55)$, $(1,0)$, and
$(1,-0.6)$. The observation time $T=5/\mbox{Im}{(z_2-\bar x)}$.
The parameters of the model are the same as in Fig.~1.}
\end{figure}

For the Zeno time $\tau_Z$, the dependence $\tau_Z$ on $\Delta$
can hardly be analyzed analytically. Nonetheless, for big values
of $\Delta$ and $N=2$ we find
 $$
 \tau_Z \approx { -\Lambda \mbox{Im}{(z_2-\bar x)} -
\log{(|\overa|^2/2)}/T \over \Lambda^2\Delta^2 (|\alpha_2|^2
-|\alpha_2|^4) + \lambda^2 \Lambda^2 F^0 |\overa|^2},
 $$
hence $\tau_Z$ decreases when $\Delta \to \infty$. We plot in
Fig.~2b the time $\tau_Z$ for the same model as for the time
$t_Z$. We can see that the ratio $t_Z/\tau_Z \approx \bar x$ is
about the same as for one-level model. The increase of the Zeno
time also takes place except for the cases when initial vector
$|\Phi\ra$ is close to the state which decays without
oscillations, i.e. $|\overa|^2=2$. This differs from the behaviour
of the Zeno time $t_Z$ where the only states, for which the
increase does not take place, are ones with one populated level:
$|\Phi\ra=(1,0)$ and $|\Phi\ra=(0,1)$. This difference is due to
the fact that while $t_Z$ is defined only by the short time
expansion, the time $\tau_Z$ depends on the behaviour of the
system in the exponential era. This dependence is also responsible
for the oscillations seen in Fig.~2b.

\section{Anti-Zeno effect}

In the previous section we have seen that for very short times $t
\lesssim t_Z (\tau_Z)$, the Zeno effect takes place. We will not
discuss here the transition region $t \sim t_Z (\tau_Z)$, where
the Zeno region changes to the anti-Zeno one, but concentrate
instead on the longer times $t \gtrsim t_Z (\tau_Z)$. We neglect
the integral term in Eq.~(\ref{samp2}), as its magnitude is
proportional to $\lambda^2$~\cite{NPN,AKPY}, and is much smaller
than effects which we discuss.
%These times correspond to the exponential era, and it is
%well-known~\cite{NPN,AKPY} that within this time interval we can
%neglect the integral term in Eq.~(\ref{samp2}).
We shall analyze the same examples as in the previous section.

{\bf Example A. The decay of one level}

The only level $l$ is initially occupied: $\alpha_l=1$,
$\alpha_k=0$ for $k \neq l$. The survival probability is here
 $$
 p(t) \sim |\sum_j r^j_{ll} e^{-i\Lambda z_j t}|^2,
 $$
and it formally coincides with the survival probability for the
one-level Friedrichs model. However, the partial resolvent $G(x)$
may be different, so the pole structure and the residues may also
be different. If the non-degenerated energy level $x_l$ is well
separated from the others, then the survival probability coincides
with one for one-level Friedrichs model. In this case, the
anti-Zeno region, defined by the value of $|r^j_{ll}|$ has already
been analyzed~\cite{AKPY,Pascazio-var}. If the energy level is
degenerate, then the situation is different as it will be
discussed in Examples B and C.

{\bf Example B. The completely degenerate case}

All energy levels are identical, $x_k= \bar x$ for any $k$.
Starting from Eq.~(\ref{ampl-deg}), we find for the survival
probability:
 \beq
p(t) \approx \left|(1 - {|\overa|^2 \over N}) +
 {|\overa|^2 \over N} e^{-i(z_2-\bar{x})\Lambda t} \right|^2
 \approx \left|(1 - {|\overa|^2 \over N}) +
 {|\overa|^2 \over N} e^{-t/t_d} e^{i s t/t_d} \right|^2,
 \eeq
where $s=\mbox{Re}{W(\bar x)}/ \mbox{Im}{W(\bar x)}$. In the last
expression we assume that $|r^j_{kl}|=1$. Then the decay rate
$\gamma(t)$ for short time can be expressed as
 \beq \label{gamma-short}
 \gamma(t) = |\overa|^2 \left( \Lambda \lambda^2 \mbox{Im} {W(\bar x)}
 + {1\over 2} t N \Lambda^2 \lambda^4
 (\mbox{Re}^2 W(\bar x)- \mbox{Im}^2 W(\bar x))\right)=
 {|\overa|^2 \over N} \left( {1\over t_d} - {t \over 2 t_d^2}(1-s^2)
 \right).
 \eeq
Expansion~(\ref{gamma-short}) can be used for the time interval
$\tau_Z \lesssim t \ll t_d$. The decay rate~(\ref{gamma-short})
should be compared with the decay rate for the exponential era:
 $$\gamma(t) \sim - \log(1-|\overa|^2/N)/t \quad \mbox{for} \quad
 |\overa|^2 \neq N. $$
If $|\overa|^2 = N$, the system completely decays without
oscillations with asymptotic decay rate $\gamma = N \Lambda
\lambda^2 \mbox{Im} W(\bar x)$ and mimics the one-level model.
Otherwise, we always have the anti-Zeno region as the variable
decay rate within the exponential era goes to zero when the time
of the observation increases.

The duration of the anti-Zeno region depends both on the initial
vector $|\Phi\ra$ and the level of degeneracy $N$. If the initial
vector $|\Phi\ra$ is such that $|\overa|^2$ is independent of the
degeneracy $N$, the decay rate for short times is not sensitive to
$N$, and the beginning of the anti-Zeno region is rather stable
with respect to $N$, see Fig.~3. However, the duration strongly
depends on $N$. We should note that in this case system decays to
different survival probabilities $p_\infty =(1-|\overa|^2/N)^2$.
If we fix the final amplitude $p_\infty$, then $|\overa|^2 = N
(1-\sqrt{p_\infty})$, and the anti-Zeno region is independent of
$N$ in the units of the decay time $t_d$, see
Eq.~(\ref{gamma-short}). Hence, the duration and the position of
the anti-Zeno region is proportional to $N$.

\begin{figure}[hbt]
 \centering \includegraphics[width=0.45\textwidth]{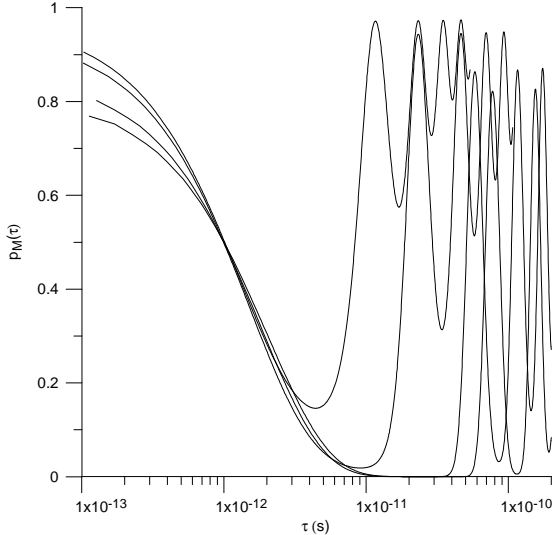}
\caption{The probability $p_M(\tau)$ for the completely
degenerated system as a function of the duration $\tau$ between
measurements. From above, the curves correspond to degeneracy
$N=20$, 10, 4, 3, and $|\overa|^2=0.9$. The observation time
$T=3/\mbox{Im}{(z_2-\bar x)}$. The parameters of the model are the
same as in Fig.~1.}
\end{figure}

{\bf Example C. $N$-level model with one different level}

In this case, all energy levels are identical except one:
$x_k=\bar{x}$, $k=1\ldots N-1$; $x_N=\bar{x}+\Delta$, and all form
factors are identical: $f_k(x)=f(x)$. The survival probability can
be calculated from Eq.~(\ref{A-1diff}). We consider the simplest
non-degenerate case with $N=2$.  The more general situation $N
\geq 3$ does not add anything essentially new. For the survival
amplitude we have
 \beq
p(t) \approx \left| {1 \over 2} e^{-i(z_2-\bar{x})\Lambda t}
|\overa|^2 + e^{-i(z_3-\bar{x})\Lambda t} (1 - {1 \over 2}
|\overa|^2 ) \right|^2.
 \eeq
The decay rate for short times is
 \beq \label{gamma-short-1diff}
 \gamma(t) = \Lambda C_1 - {\Lambda^2 |\overa|^2 C_2 \over 4} t +O(t^2),
 \eeq
where
 $$ C_1= - {|\overa|^2 \over 2} \mbox{Im} z_2 - (1-{|\overa|^2
 \over 2}) \mbox{Im} z_3 >0, \quad C_2 = (1+{|\overa|^2 \over 2})
 (\mbox{Im}(z_2-z_3))^2 - (1-{|\overa|^2 \over 2})
 (\mbox{Re}(z_2-z_3))^2. $$
For the case $\Delta \ll \lambda^2$ and for long times we find
$\gamma(t) \sim - \mbox{Im} z_3$. Therefore, for the initial
vector $|\Phi\ra$ such that $|\overa|^2 \neq 0$, the decay rate
for short times is bigger than $-\mbox{Im} z_3$. In this case a
wide stable anti-Zeno region of the size $\sim \mbox{Im} z_2$
appears (see Fig.~4). When the $\Delta$ becomes bigger, the
position and the size of the anti-Zeno region considerably
changes.

\begin{figure}[hbt]
 \centering \includegraphics[width=0.45\textwidth]{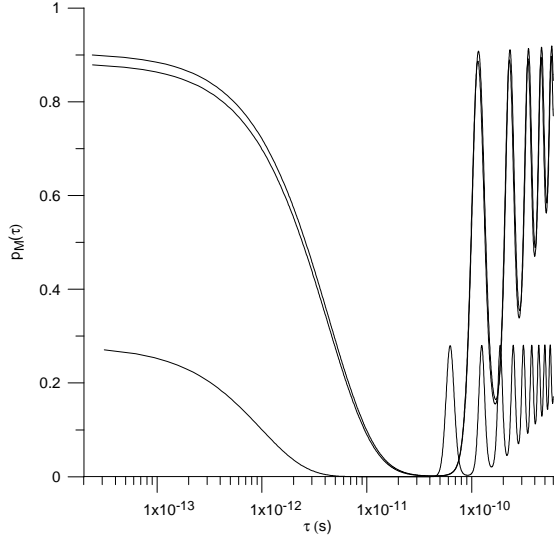}
\caption{The probability $p_M(t)$ for the model with two close
levels as a function of the duration $\tau$ between measurements.
From above, the curves correspond to energy splitting $\Delta=
10^{-12}$, $10^{-9}$, $10^{-8}$, and $|\overa|^2=0.05$. The
observation time $T=2/\mbox{Im}{(z_2-\bar x)}$. The parameters of
the model are the same as in Fig.~1.}
\end{figure}

The origin of the anti-Zeno behaviour for Example~A and
Examples~B, C is very different. For Example~A, as well for the
one-level Friedrichs model, the anti-Zeno behaviour is due to the
fact the the residues $r_{ll}^j$ are not equal to $1$:
$|r_{ll}^j-1| \sim \lambda^2$. While this is also true for
Examples~B and C, this difference is much smaller than the
non-exponentiality due to the degeneracy (or nearly degeneracy) of
the energy levels. This non-exponentiality is the origin of the
anti-Zeno behaviour in Examples~B and C.

\section{Concluding remarks}

We have considered the temporal behaviour of the survival
probability of excited states in the $N$-level Friedrichs model
for degenerate and nearly degenerate situations. For various
initial conditions we have determined the duration of the Zeno and
anti-Zeno eras and analyzed the behaviour on the intermediate
exponential era where we have found a rich variety of behaviour
from pure exponential decay to exponentially decaying
oscillations. For initial states belonging to the nondecaying
subspaces, these oscillations stabilize without decaying to zero.
The experimental implementation of this result should be exploited
as a mean of suppression of decoherence.

In the short-time scale, our analysis has shown also a possibility
for considerable slowing down of the decay due to the Zeno effect
in the nearly degenerate system for a special class of initial
conditions. If these systems and conditions are realizable
experimentally(in atoms, ions, quantum dots etc.), one has a new
possibility for efficient suppression of decoherence in quantum
computation and communication using the Zeno effect. We have
analyzed and compared two different definitions of the Zeno time.
This allows for a better estimation of the time intervals where
the Zeno effect may be really helpful in quantum computations.

\acknowledgments
 We would like to thank Prof. Ilya Prigogine for helpful discussions.
This work was supported by the European Commission Project No.
HPHA-CT-2001-40002 (NSNS).

\end{document}